\documentclass[aps,prl,twocolumn,superscriptaddress, floatfix]{revtex4}
\usepackage{amsbsy,amssymb,amsmath,bm}
\usepackage{graphicx,color,rotate}
\usepackage{units}
\usepackage{floatrow}
\usepackage{multirow}

\begin{document}

\title{Bulk Fermi surface of the Weyl type-II semimetallic candidate NbIrTe$_{4}$}

\author{Rico Sch{\"o}nemann}\email{schoenemann@magnet.fsu.edu}
\affiliation{National High Magnetic Field Laboratory, Florida State University, Tallahassee, Florida 32306, USA}
\author{Yu-Che Chiu}
\affiliation{National High Magnetic Field Laboratory, Florida State University, Tallahassee, Florida 32306, USA}
\affiliation{Department of Physics, Florida State University, Tallahassee, Florida 32306, USA}
\author{Wenkai Zheng}
\affiliation{National High Magnetic Field Laboratory, Florida State University, Tallahassee, Florida 32306, USA}
\author{Victor Quito}
\affiliation{Department of Physics and Astronomy, Iowa State University, Ames, Iowa 50011, USA}
\author{Shouvik Sur}
\affiliation{Department of Physics and Astronomy, Northwestern University, Evanston, Illinois 60208, USA}
\affiliation{Department of Physics, Florida State University, Tallahassee, Florida 32306, USA}
\author{Gregory T. McCandless}
\affiliation{Department of Chemistry and Biochemistry, The University of Texas at Dallas, Richardson, Texas 75080 USA}
\author{Julia Y. Chan}
\affiliation{Department of Chemistry and Biochemistry, The University of Texas at Dallas, Richardson, Texas 75080 USA}
\author{Luis Balicas}\email{balicas@magnet.fsu.edu}
\affiliation{National High Magnetic Field Laboratory, Florida State University, Tallahassee, Florida 32306, USA}

\date{\today}

\begin{abstract}

Recently, a new group of layered transition-metal tetra-chalcogenides was proposed via first principles calculations to correspond to a new family of Weyl type-II semimetals with promising topological properties in the bulk as well as in the monolayer limit. In this article, we present measurements of the Shubnikov-de Haas (SdH) and de Haas-van Alphen effects under high magnetic fields for the type-II Weyl semimetallic candidate NbIrTe$_{4}$. We find that the angular dependence of the observed Fermi surface extremal cross-sectional areas agree well with our DFT calculations supporting the existence of Weyl type-II points in this material. Although we observe a large and non-saturating magnetoresistivity in NbIrTe$_{4}$ under fields all the way up to $35\,\mathrm{T}$, Hall-effect measurements indicate that NbIrTe$_{4}$ is not a compensated semimetal.
The transverse magnetoresistivity displays a four-fold angular dependence akin to the so-called butterfly magnetoresistivity observed in nodal line semimetals.
We conclude that the field and this unconventional angular-dependence are governed by the topography of the Fermi-surface and the resulting anisotropy in effective masses and in carrier mobilities.

\end{abstract}

\maketitle

\section{Introduction}

Recently, Weyl fermions have emerged as a heavily studied subject combining key concepts from high energy and condensed matter physics \cite{reviews}.
In a Weyl semimetal, Weyl fermions emerge around the touching points between linearly dispersing valence and conduction bands.
Type-I Weyl points correspond to the ``conventional" Weyl Fermions of quantum field theory. Weyl semimetals require broken inversion or time reversal symmetry for the Weyl nodes of opposite chirality to separate in $k$-space or to prevent their pairwise annihilation. Insofar, several candidates for type-I Weyl semimetals have been found, most notably the compounds belonging to the TaAs family \cite{huang_weyl_2015, xu_weyl_2015, lv_experimental_2015, weng_weyl_2015}. In these compounds angle resolved photoemission spectroscopy (ARPES) and magneto transport experiments were able to reveal a few characteristic signatures of Weyl semimetals like topological Fermi arcs on their surface, the emergence of a controversial negative longitudinal magnetoresistance (NLMR) when magnetic and electric fields are aligned \cite{huang_observation_2015}, and the observation of the so-called Weyl-orbits  \cite{moll}.  exploring the Fermi arcs.

Weyl type-II fermions are predicted to emerge at the boundary between hole- and electron-pockets resulting from strongly tilted Weyl cones that break Lorentz invariance within the crystal and, therefore, have no equivalents in high energy physics \cite{soluyanov_type-ii_2015}. Candidates for Weyl type-II semimetals include non-centrosymmetric materials whose inversion symmetry is broken like the layered orthorhombic transition-metal dichalcogenides MoTe$_{2}$ and WTe$_{2}$ \cite{sun_prediction_2015, wang_mote_2016}, as well as MoP$_{2}$ and WP$_{2}$ \cite{autes_robust_2016} which do not crystallize in a layered structure. However, type-II weyl points have been predicted to emerge in the ternary tellurides TaIrTe$_{4}$ and NbIrTe$_{4}$ \cite{koepernik_tairte_2016, li_ternary_2017}. In fact, the entire family of the $MM'$Te$_{4}$, where $M$ = Ta or Nb and $M'$ = Ir or Rh, is candidate for bulk Weyl type-II semimetallic states and is also predicted to display a topological quantum spin hall insulating phase in the monolayer limit \cite{liu_van_2017}. TaIrTe$_{4}$ was the first representative within this family of materials to be claimed to exhibit a Weyl semimetallic state \cite{koepernik_tairte_2016}. It was shown that by including spin-orbit coupling (SOC), TaIrTe$_{4}$ hosts a minimum of four Weyl points within the first Brillouin zone. Experimental results from quantum oscillations \citep{khim_magnetotransport_2016} and ARPES \citep{belopolski_signatures_2017, haubold_experimental_2017} measurements where able to identify the Weyl points by comparing the measured electronic structure with the DFT calculations. Interestingly TaIrTe$_{4}$ displays a complex electronic structure that can be tuned by external factors like strain which modifies the number of Weyl points or the location of topological nodal lines \cite{zhou_coexistence_2018}.

In this article, we investigate the topography of the Fermi surface of the Weyl type-II semimetallic candidate NbIrTe$_{4}$ via measurements of the de Haas-van Alphen (dHvA) and the Shubnikov-de Haas (SdH) effects to compare with band structure calculations. Similar to $T_d$-MoTe$_{2}$ and WTe$_{2}$, NbIrTe$_{4}$ is a non-centrosymmetric layered compound belonging to the orthorhombic space group \textit{Pmn}2$_{1}$ \cite{mar_metal-metal_1992}, as shown in Fig. 1 (a). Based on DFT calulations, eight Weyl points emerge within the first Brillouin zone in the absence of SOC. After including SOC a total of 16 Weyl points emerge between the topmost valence and the lowest conduction band, which makes the electronic structure of NbIrTe$_{4}$ more complex than that of its sister compound TaIrTe$_{4}$. These nodes are located within $142\,\mathrm{meV}$ of the Fermi energy $E_{\mathrm{F}}$ with 8 nodes located in the $k_z =0$ plane and other 8 in $k_z = \pm 0.2$ planes \cite{li_ternary_2017}.

\section{Methods}

\begin{figure}
\begin{center}
		\includegraphics[width = 8.6 cm]{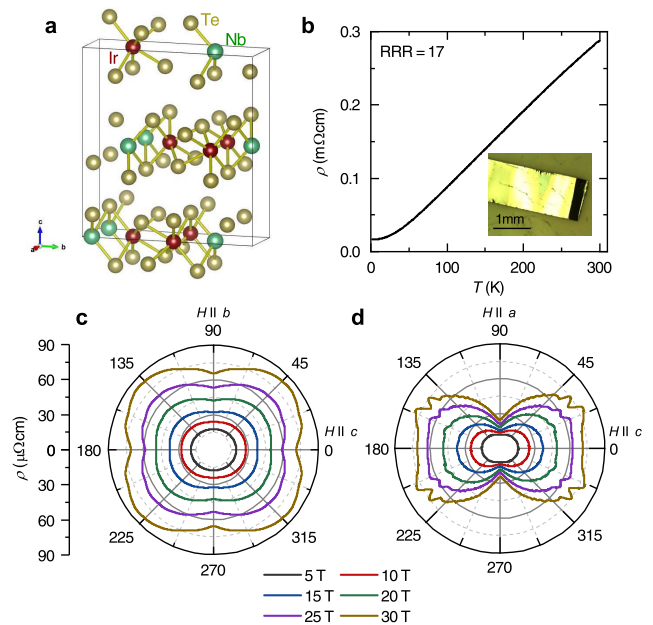}
		\caption{(a) Crystallographic structure of NbIrTe$_{4}$. Te atoms are depicted as golden spheres while Ir and Nb atoms are represented by red and green spheres, respectively. (b) Resistivity as a function of the temperature for a NbIrTe$_{4}$ single-crystal. The inset shows a picture of a typical NbIrTe$_{4}$ crystal. (c) and (d) Angular dependence of the resistivity under different magnetic fields at $T = 0.35$ K. In (c) the field was rotated within the $bc$-plane of the crystal with the current $I\parallel a$, $0^{\circ}$ corresponds to $\mu_0H\parallel c$ and $90^{\circ}$ to $\mu_0H\parallel b$. In panel (d) the field was rotated within the $ac$-plane with $I\parallel b$, $0^{\circ}$ is $\mu_0 H\parallel c$ and $90^{\circ}$ is $\mu_0H\parallel a$.}
	\label{fig:Struct_RT_Ang}
\end{center}
\end{figure}

Single crystals of NbIrTe$_{4}$ were grown via a Te flux method. Stoichiometric amounts of elementary Nb ($99.999\%$ Alfa Aesar) and Ir ($99.99\%$ Sigma Aldrich) with excess Te were heated in a sealed quartz ampule up to $1000^{\circ}\mathrm{C}$ and then slowly cooled to  $700^{\circ}\mathrm{C}$. After removing the excess Te via centrifugation of the ampoules at  $700^{\circ}\mathrm{C}$, we obtained shiny metallic crystals with dimensions up to $5 \times 1 \times 0.1$ mm$^{3}$. The inset in Fig. \ref{fig:Struct_RT_Ang}(b) shows an image of a typical bar-shaped NbIrTe$_{4}$ single-crystal. The crystallographic $a$-axis is in general aligned along the longest crystal dimension and the $c$-axis along the shortest. The composition and phase purity of our samples was confirmed by energy-dispersive X-ray spectroscopy (EDS) and X-ray diffraction. Initial resistivity and Hall-effect measurements on NbIrTe$_{4}$ were performed in a $^{4}\mathrm{He}$-cryostat equipped with a $9\,\mathrm{T}$ superconducting magnet (Quantum Design PPMS). Experiments under high magnetic fields up to $35\,\mathrm{T}$ where performed in resistive Bitter magnets at the National High Magnetic Field Laboratory (NHMFL) in Tallahassee using $^{3}$He cryostats. AC resistivity measurements were performed on single-crystals using the standard 4-wire technique. Additionally, a capacitive cantilever beam technique was used for magnetic torque measurements.

\begin{figure}
\begin{center}
		\includegraphics[width = 8.6 cm]{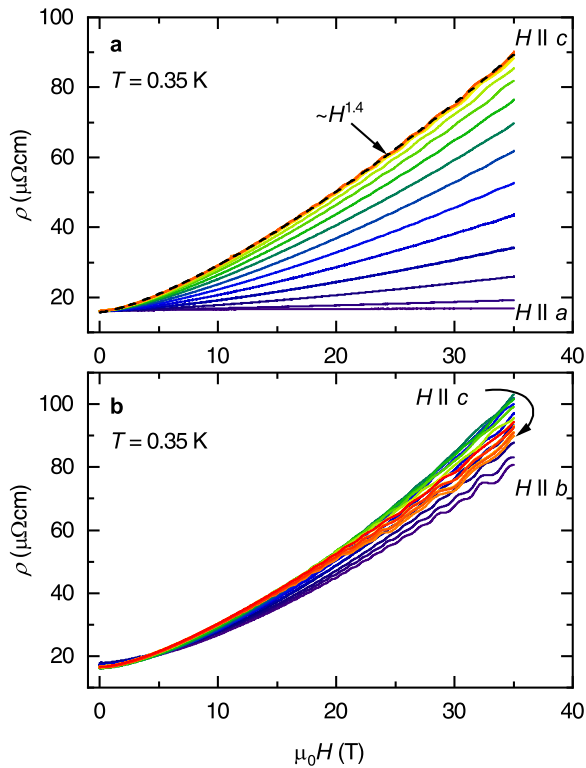}
		\caption{(a) Resistivity $\rho$ as a function of the magnetic field $\mu_0H$ for different angles between $\mu_0 H\parallel c$ and $\mu_0 H\parallel a$-axis. Black dashed line depicts a power law fit of $\rho(\mu_0 H)$ for $\mu_0 H\parallel c$ (red curve). (b) $\rho$ as a function of $\mu_0 H$ for different field orientations in the $bc$-plane. Above $\mu_0H \sim 10$ T quantum oscillations superimposed onto the magnetoresistive background are clearly visible in both panels.}
	\label{fig:MRAng}
\end{center}
\end{figure}

In order to obtain the electronic band structure of NbIrTe$_{4}$ and the geometry of its Fermi surface we performed DFT calculations including spin-orbit coupling using the Wien2k package \cite{blaha_full-potential_1990}. The Perdew-Burke-Ernzerhof (PBE) exchange correlation functional \cite{perdew_generalized_1996} was used in combination with a dense $k$-mesh of $22 \times 8 \times 8$ $k$-points and a cutoff $RK_{\mathrm{max}}$ of 7.5. The structural parameters were taken from Ref. \cite{mar_metal-metal_1992}. The angular dependence of the SdH and dHvA frequencies, which are associated to the extremal cross-sectional areas of the Fermi surface through the Onsager relation, the effective cyclotron masses and the charge carrier densities were calculated using the SKEAF code \cite{rourke_numerical_2012}. For the visualization of the crystal structure and of Fermi surface we used the XCrysden \cite{kokalj_xcrysdennew_1999} package.

\begin{figure*}
\begin{center}
		\includegraphics[width = 15 cm]{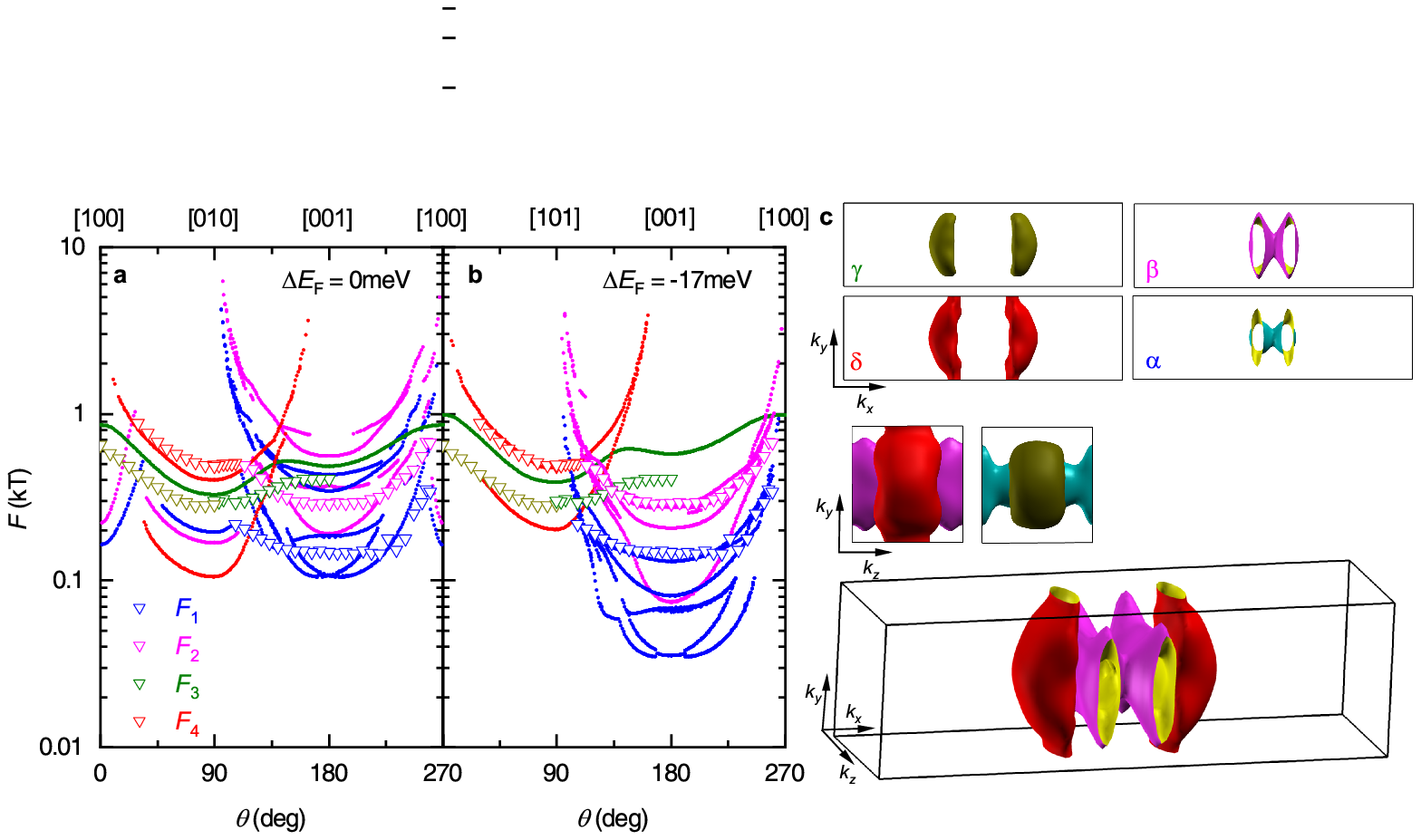}
		\caption{(a) and (b) Angular dependence of the SdH frequencies $F(\theta)$ for NbIrTe$_{4}$.  (a) $F(\theta)$ for the position of the Fermi energy $E_{\mathrm{F}}$ resulting from the DFT calculations. (b) $F(\theta)$ with $E_{\mathrm{F}}$ shifted by $-17\,\mathrm{meV}$. The smaller solid points represent the SdH frequencies obtained from DFT calculations using the Onsager relation. Larger triangles depict the position of the peaks observed in the Fourier transform of the experimental oscillatory signal superimposed onto the resistivity data. Notice the better agreement between the experimental data and the frequencies resulting from the shift of the Fermi level.  $F_{3}$ and $F_{4}$ can be assigned to the hole pockets labeled as $\gamma$ and $\delta$ in (c), $F_{1}$ and $F_{2}$ can be assigned to the electron pockets $\alpha$ and $\beta$.}
	\label{fig:SdHAngDep}
\end{center}
\end{figure*}

\section{Results}

The layered structure of NbIrTe$_{4}$ is displayed in Fig. \ref{fig:Struct_RT_Ang}(a) and it corresponds to a variant of the WTe$_{2}$ one as discussed in Ref. \cite{mar_metal-metal_1992}. Nb and Ir atoms form zigzag chains along the $a$-direction. When compared to WTe$_{2}$ the alternation of Nb and Ir results in a doubling of the unit cell along the $b$-axis. Measurements of the resistivity $\rho$ in the absence of an external magnetic field reveal metallic behavior with $\rho(T)$ saturating at low temperatures around a residual resistivity $\rho_{0} \simeq 16$ $\mu\Omega$cm (see, Fig. \ref{fig:Struct_RT_Ang}(b)). The residual resistivity ratio (RRR) which is defined here as $\mathrm{RRR} = \rho(300 {\text{ K}})/\rho_{0}$ reaches a value of 17 for this particular sample. However, we consistently found RRR values ranging from 15 to 40 across several sample batches.  Although these RRR values are not particularly high, when compared to other layered transition metal chalcogenides like MoTe$_{2}$, the resulting residual resistivites are relatively low with the presence of quantum oscillations confirming that these crystals are of relative high quality. In this manuscript, we include data from 4 NbIrTe$_{4}$ single-crystals used for resistivity and torque measurements at high fields.

The angular dependence of the transverse magnetoresistivity $\rho(\theta)$ of NbIrTe$_{4}$ for fields rotating within the $ac$-plane, and for $I \parallel b$-axis, is shown in Fig. \ref{fig:Struct_RT_Ang}(d). $\rho(\theta)$ is two-fold symmetric with the additional structure emerging at higher fields resulting from the SdH-effect. Its minimum is observed for fields parallel to the $a$-axis, implying increased inter-layer scattering and a smaller carrier mobility and larger effective masses for fields along this orientation.

For fields rotating in the $bc$-plane, as in shown in Fig. \ref{fig:Struct_RT_Ang}(c), $\rho(\theta)$ displays a significantly smaller anisotropy, when compared to fields rotating in the $ac$-plane, reaching its maximum for $\theta \approx 45^{\circ}$ and its minimum for $H\parallel b$. This anisotropy results in a four-fold symmetric ``butterfly" shaped angular dependence that is only present under magnetic fields exceeding $\mu_0H = 10\,T$. Under smaller fields the butterfly disappears and $\rho(\theta)$ becomes maximal for $H\parallel c$. This butterfly shaped magnetoresistance was also  observed in the high-$T_{\mathrm{c}}$ superconductors \cite{raffy}, in magnetic thin films, as well as in the Dirac nodal line semimetal ZrSiS \cite{ali_butterfly_2016}. In the case of ZrSiS this behavior was ascribed to a topological phase-transition as a function of field orientation that is inherent to the nodal Dirac line  \cite{ali_butterfly_2016} although the SOC should gap the nodal lines that are located in close proximity to its Fermi level given that they are associated to symmorphism.

Similarly to TaIrTe$_{4}$ \cite{khim_magnetotransport_2016}, the negative longitudinal magnetoresistance (NLMR) observed in the Weyl type-I monopnictide semimetals \cite{huang_observation_2015, hu__2016, arnold_negative_2016}, which was ascribed to the axial anomaly between Weyl points, is absent in NbIrTe$_{4}$ for fields and currents along the $a$-axis. For Weyl type-II semimetals the positive longitudinal magneto-conductivity observed in Weyl type-I was originally predicted to depend on the orientation of the external magnetic field relative to wave-vector connecting the Weyl nodes \cite{soluyanov_type-ii_2015, udagawa}. Although, more recently it was claimed to be orientation independent \cite{pallab}. In NbIrTe$_{4}$, as well as in TaIrTe$_{4}$, perhaps the Weyl nodes are located too far away from the Fermi level to lead to charge carriers having a well-defined chirality or perhaps that their electrical transport properties are dominated by the topologically trivial bands.

Figure \ref{fig:MRAng} displays the magneto-resistivity $\rho(\mu_0H)$ as a function of the angle $\theta$ for fields rotating within the $ac$ and the $bc$ planes. For $\mu_0H\parallel c$-axis SdH oscillations, superimposed onto the magnetoresistive background, become observable when the field exceeds $\mu_0 H \sim 8$ T. But their amplitude weakens as the field is oriented towards the $a$-direction, limiting the observation of quantum oscillations to $\theta \lesssim 70^{\circ}$ with respect to the $c$-axis. For rotations within the $bc$-plane the SdH oscillations are visible over the entire angular range. Remarkably, over a decade in field the magnetoresistivity can be fit to a single power law where $\rho(\mu_0H) \propto H^{\alpha}$ yielding $\alpha=1.42$. Figure S1 in the Supplementary Information file \cite{supplemental} displays the magnetoresistivity in a log-log scale indicating quite clearly the existence of a single, anomalous power law over an extended range of magnetic fields with this exponent being nearly orientation independent. Notice that this value is quite close to the value $\alpha \simeq 1.5$ extracted for TaIrTe$_4$ \cite{khim_magnetotransport_2016}. The first step to address this behavior would be to develop a model within Boltzmann transport theory combining both closed and open orbits, given the geometry of the Fermi surface derived from the calculations which, as shown below, is confirmed by our experiments, including anisotropic effective masses. As discussed by Ref. \cite{mr_fermi}, such approach is capable of describing non-saturating magnetoresistivity displaying an unconventional power dependence on field in addition to replicating its anisotropy. If conventional transport theory was unable to capture this behavior in a scenario that includes the lack of carrier compensation, one could conjecture that it might bear relation to the existence unconventional quasiparticles.

\begin{figure}
\begin{center}
		\includegraphics[width = 8.6 cm]{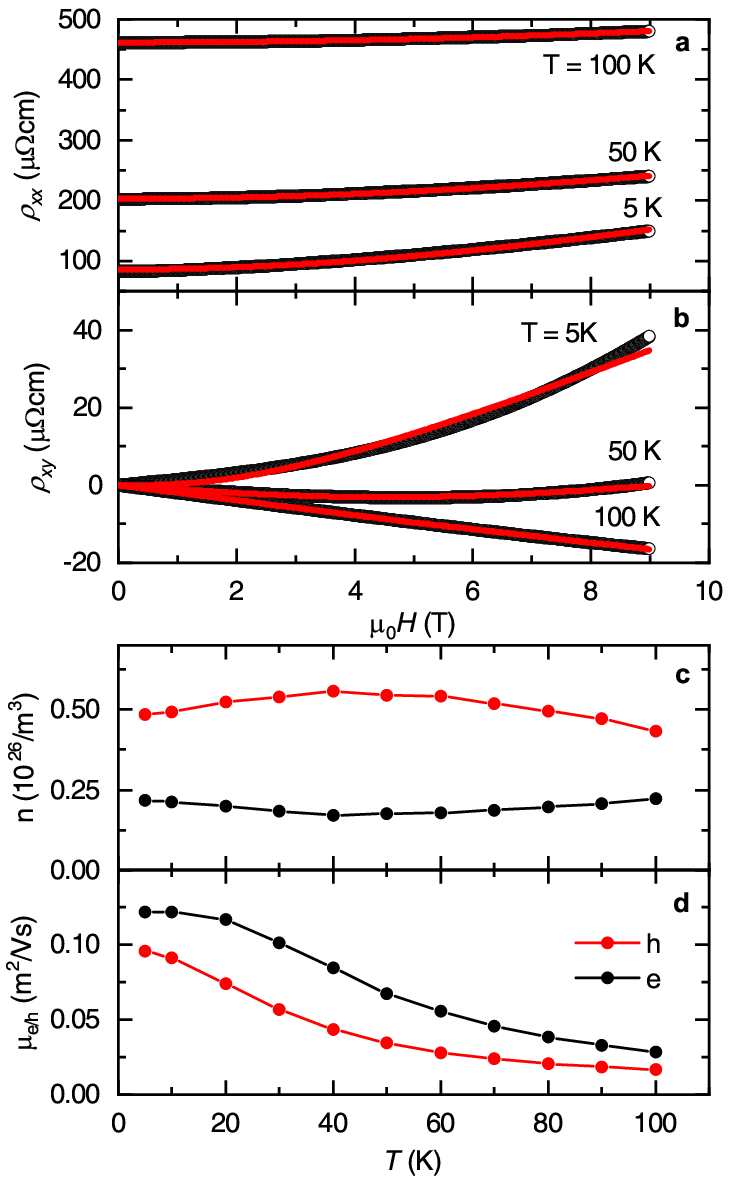}
		\caption{(a, b) Longitudinal resistivity $\rho_{xx}$ and Hall resistivity $\rho_{xy}$ for NbIrTe$_{4}$ as a function of the magnetic field for $T = 5\,\mathrm{K}$, $50\,\mathrm{K}$ and $100\,\mathrm{K}$. The red lines are fits of $\rho_{xx}$ and $\rho_{xy}$ to the two band model (see equation \ref{eq:2band}).(c, d) Electron and hole carrier densities $n_{\mathrm{e/h}}$ and carrier mobilities $\mu_{\mathrm{e/h}}$ as a function of the temperature $T$. $n_{\mathrm{e/h}}$ and $\mu_{\mathrm{e/h}}$ have been extracted from simultaneous fits of the Hall-effect and of the magnetoresistivity data to the two-band model}
	\label{fig:HallvsT}
\end{center}
\end{figure}

\begin{figure}
\begin{center}
		\includegraphics[width = 8.6 cm]{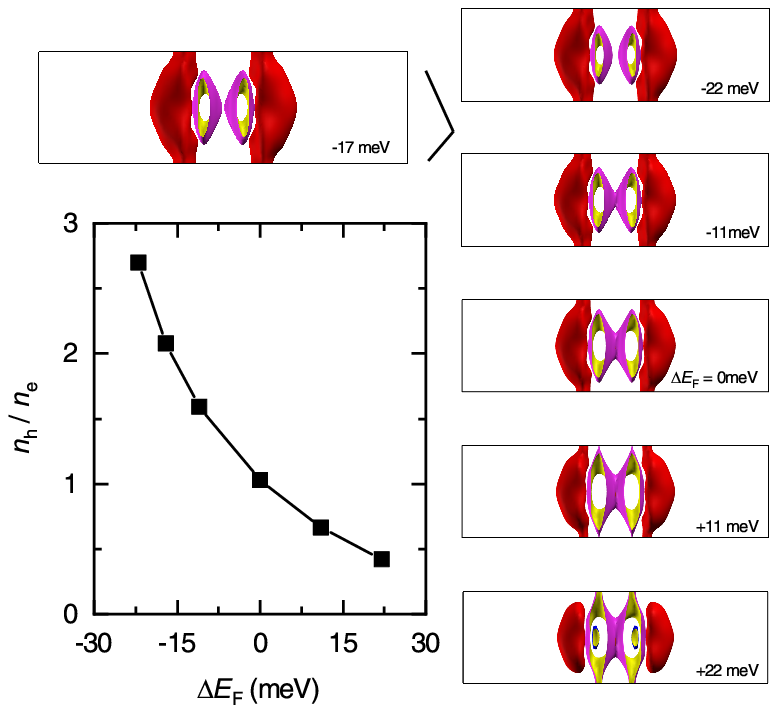}
		\caption{Fermi surface of NbIrTe$_{4}$ projected on the $k_{y}k_{z}$-plane for different Fermi energies. The graph shows the calculated $n_{\mathrm{h}}/n_{\mathrm{e}}$ ratio as a function of the shift in Fermi energy $\Delta E_{\mathrm{F}}$.}
	\label{fig:CarrierDens}
\end{center}
\end{figure}

\begin{figure*}
\begin{center}
		\includegraphics[width = 15 cm]{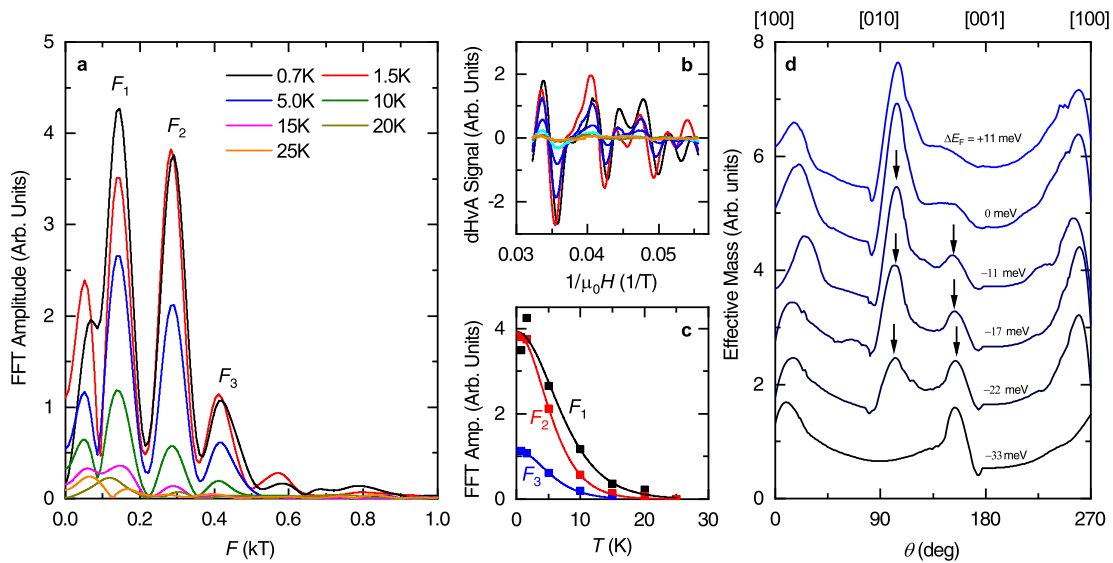}
		\caption{Effective masses of NbIrTe$_{4}$ extracted from quantum oscillations. (a) Fast Fourier Transform (FFT) of the oscillatory component on the torque signal shown in (b) for $H\parallel c$. Frequencies that can be assigned to individual orbits on the Fermi surface pockets are labeled as $F_{1}$, $F_{2}$ and $F_{3}$. (b) de Haas-van Alphen oscillations in NbIrTe$_{4}$ for temperatures between $0.7\,\mathrm{K}$ and $T = 20\,\mathrm{K}$. (c) FFT amplitude as a function of the temperature for the dHvA frequencies $F_{1}$, $F_{2}$ and $F_{3}$. Solid lines represent fits to the temperature damping factor $R_{T}$ in the Lifshitz-Kosevich formalism from which we extract the effective masses. (d) Angular dependence of the cyclotron effective mass averaged across all extremal orbits as obtained from the DFT calculations. For the sake of clarity, these curves are shifted with respect to their respective Fermi levels. The curve assigned to $\Delta E_{\mathrm{F}} = 0$ corresponds to the unshifted Fermi energy. Arrows indicate maxima in the average effective mass for fields oriented nearly along the $b$- or the $c$-axis. We argue that this anisotropy in effective masses leads to the butterfly shaped magnetoresistivity.}
	\label{fig:TorqueEffMass}
\end{center}
\end{figure*}

In order to reveal the topography of the Fermi surface of NbIrTe$_{4}$, we performed magneto-resistivity measurements as a function of the angle $\theta$ to extract the angular dependence of the SdH frequencies. The results are summarized in Fig. \ref{fig:SdHAngDep}, see also Fig. S3 in SI \cite{supplemental}. To obtain the oscillatory signal, we fit the magnetoresistive background to a polynomial and subsequently subtract it from the experimental curve. The fast Fourier transform of the oscillatory component superimposed onto the resistivity and torque data can be found in the SI material (see, Fig. S3). According to the DFT calculations the Fermi surface of NbIrTe$_{4}$ consists of two pairs of spin-orbit split electron (labeled as $\gamma$ and $\delta$) and hole-pockets ($\alpha$ and $\beta$) located near the center of the Brillouin Zone (see, Fig. \ref{fig:SdHAngDep}(c)). Three of the four pockets ($\alpha$, $\beta$, and $\delta$) are strongly corrugated cylinders aligned along the $k_{z}$ and the $k_{y}$-directions, respectively. The $\gamma$ sheet forms an anisotropic ``kidney" shaped pocket. As shown in Fig. \ref{fig:TorqueEffMass}, for magnetic fields oriented along the $c$-axis we can identify three distinct frequencies $F_{1}$, $F_{2}$ and $F_{3}$. Based on their angular dependence $F_{1}$ and $F_{2}$ can be assigned to the electron pockets and $F_{3}$ to the kidney shaped hole-pocket. Although the lower calculated frequencies associated to the hole pockets are not clearly visible in the experimental data, we can achieve a quite acceptable agreement between calculated and experimental frequencies by lowering the Fermi level by $-17\,\mathrm{meV}$. Only the size of the $\gamma$-pocket is overestimated in the calculations by approximately $30\%$. From the evolution of the Fermi surface with respect to the position of the Fermi energy it is evident that a shift of the $\gamma$ pocket does not affect the Weyl points in NbIrTe$_{4}$ since they appear at touching points between the $\delta$ and the $\beta$ pockets. Since the Weyl type-II points result from band crossings not associated with the band yielding the $\gamma$ sheet, one can safely state that these nodes are not affected by the accurate position of this band relative to the Fermi level, see band depicted by green line in Fig. S4 within the SI file \cite{supplemental}.

Further justification for lowering the Fermi level comes from Hall-effect measurements that were performed on a mechanically exfoliated sample with a thickness of $20\,\mathrm{\mu m}$ and lateral dimensions of 1-$2\,\mathrm{mm}$. We extract the charge carrier densities and mobilities of NbIrTe$_{4}$ from Hall-effect measurements collected between 5 and $100\,\mathrm{K}$ under magnetic fields up to 9 T ($\mu_0H\parallel c$-axis) by simultaneously fitting the longitudinal magneto-resistivity $\rho_{xx}$ and the Hall resistivity $\rho_{xy}$ to the two-band model:

\begin{align}
\rho_{xx} & = \frac{1}{e}\frac{\left(n_{\mathrm{h}}\mu_{\mathrm{h}} + n_{\mathrm{e}}\mu_{\mathrm{e}}\right) + \left(n_{\mathrm{h}}\mu_{\mathrm{e}} + n_{\mathrm{e}}\mu_{\mathrm{h}}\right)\mu_{\mathrm{h}}\mu_{\mathrm{e}}B^{2}}{\left(n_{\mathrm{h}}\mu_{\mathrm{h}} + n_{\mathrm{e}}\mu_{\mathrm{e}}\right)^{2} + \left(n_{\mathrm{h}} - n_{\mathrm{e}}\right)^{2}\mu_{\mathrm{h}}^{2}\mu_{\mathrm{e}}^{2}B^{2}}\\
\rho_{xy} & = \frac{B}{e}\frac{\left(n_{\mathrm{h}}\mu_{\mathrm{h}}^{2} - n_{\mathrm{e}}\mu_{\mathrm{e}}^{2}\right) + \left(n_{\mathrm{h}} - n_{\mathrm{e}}\right)\mu_{\mathrm{h}}^{2}\mu_{\mathrm{e}}^{2}B^{2}}{\left(n_{\mathrm{h}}\mu_{\mathrm{h}} + n_{\mathrm{e}}\mu_{\mathrm{e}}\right)^{2} + \left(n_{\mathrm{h}} - n_{\mathrm{e}}\right)^{2}\mu_{\mathrm{h}}^{2}\mu_{\mathrm{e}}^{2}B^{2}}\label{eq:2band}
\end{align}

\noindent where $n_{\mathrm{e}}$, $n_{\mathrm{h}}$ are the electron and hole carrier densities and $\mu_{\mathrm{e}}$, $\mu_{\mathrm{h}}$ are the electron and hole carrier mobilities, respectively. In this field interval one obtains a reasonable agreement between the experimental data and the fittings, see Fig. \ref{fig:HallvsT}(a, b). The mobilities increase as the temperature is lowered to $\mu_{\mathrm{e}} \approx  0.12 \times 10^{4}\,\mathrm{cm^{2}/Vs}$ and $\mu_{\mathrm{h}} \approx 0.1 \times 10^{4}\,\mathrm{cm^{2}/Vs}$ at $T = 5\,\mathrm{K}$, while the electron and the hole densities vary little within this temperature interval. From the low temperature moblities we can estimate the classical transport lifetime $\tau_{\mathrm{D}}$ given by the Drude model, $\tau_{\mathrm{D}} = \mu_{\mathrm{e/h}}m^{*}/e = 3.5\times 10^{-13}\,\mathrm{s}$ with $\mu_{\mathrm{e/h}} \approx 0.1\,\mathrm{m^{2}/Vs}$ and an effective mass of $m^{*} = 0.5m_{\mathrm{e}}$. This result is comparable with the quantum lifetime obtained from SdH oscillations that is related to the Dingle temperature $T_{\mathrm{D}}$: $\tau_{Q} = \hbar / 2\pi k_{\mathrm{B}}T_{\mathrm{D}} = 2.3\times 10^{-13}\,\mathrm{s}$. As shown in Fig. \ref{fig:HallvsT}(d), $\mu_{\mathrm{h}}$ is smaller than $\mu_{e}$ for fields parallel to $c$ which can be attributed to open Fermi surface pockets that result in larger effective masses for the hole orbits when compared to the closed electron orbits. 

At a temperature of $5\,\mathrm{K}$ we obtain $n_{\mathrm{e}}\approx 0.22 \times 10^{20}\,\mathrm{cm^{-3}}$ and $n_{\mathrm{h}}\approx 0.48 \times 10^{20}\,\mathrm{cm^{-3}}$. Therefore, the density of holes exceeds the density of electrons by a factor greater than 2. In contrast, the volumes of the individual Fermi surface sheets for the unshifted Fermi level (Fig. \ref{fig:CarrierDens}), would yield nearly equal electron and hole densities, making NbIrTe$_{4}$ a compensated semimetal. As depicted in Fig. \ref{fig:CarrierDens} lowering the Fermi level leads to an expansion of the volume of the hole pockets while shrinking the electron pockets without fundamentally changing their shape. Thus lowering the Fermi level leads to a reduction in $n_{\mathrm{e}}$ and to an increase in $n_{\mathrm{h}}$ resulting in a better agreement with the Hall-effect data. Hence, NbIrTe$_4$ is remarkable for not being carrier compensated, and this is consistent with its modest magnetoresistivity, i.e. $\Delta\rho \sim 400$\% observed at low $T$s under fields up to $\mu_0H = 35$ T, while displaying non-saturating magnetoresistivity.

The cyclotron effective mass of a given electronic orbit can be extracted from the temperature damping factor $R_{T}$ in the Lifshitz-Kosevich formula and is given by: $R_{T} = \lambda T / \sinh(\lambda T)$ where $\lambda = 2\pi^{2}k_{\mathrm{B}}m^{*}/\hbar eB$, with $m^{*}$ being the cyclotron effective mass. The fast Fourier transform (FFT) of the oscillatory component superimposed onto the torque data for $\mu_0H$ nearly parallel to the $c$-axis is shown in the Fig. \ref{fig:TorqueEffMass}(a). There are at least three distinct frequencies, which we label as $F_{1}$, $F_{2}$ and $F_{3}$. Each correspond to a distinct extremal cross-sectional area of the Fermi surface and belong to a different Fermi surface pocket ($\alpha$, $\beta$, $\gamma$). To extract $m^{*}$ for each individual orbit, we fit $R_{T}$ to the temperature dependence of the amplitude of the peaks observed in the FFT spectra (Fig. \ref{fig:TorqueEffMass}(b)). As shown in Table \ref{tab:effmass}, $m^{*}$ is anisotropic or depends on sample orientation with respect to $\mu_0H$ ranging from 0.46 to twice the free electron mass. These values are comparable to those extracted for TaIrTe$_{4}$ \cite{khim_magnetotransport_2016}.

{
\setlength{\tabcolsep}{10pt}
\renewcommand{\arraystretch}{1.5}
\begin{table}[H]
   \caption{Effective masses of NbIrTe$_{4}$ for selected SdH and dHvA frequencies. The pockets are labeled as $\alpha$, $\beta$, $\gamma$ and $\delta$ following Figs. (\ref{fig:SdHAngDep}) and (\ref{fig:TorqueEffMass}). The angle $\theta$ represents the orientation of the magnetic field where $\theta = 0{}^{\circ} = 270{}^{\circ}\equiv \mu_0H\parallel a$, $90{}^{\circ}\equiv \mu_0H\parallel b$ and $180{}^{\circ}\equiv \mu_0H\parallel c$, where $m^{*}$ is the cyclotron effective mass in units of the free electron mass $m_{0}$, and $F$ the SdH/dHvA frequency.}
   \label{tab:effmass}
   \centering 
   \begin{tabular}{llll}
   \hline
   pocket & $\theta\,({}^{\circ})$ & $m^{*}\,(m_{0})$ & $F\,(\mathrm{kT})$ \\
   \hline
   \multirow{2}{*}{$\alpha$} & 135 	& 0.76 & 0.178\\
                             & 180 ($c$) & 0.47 & 0.148 \\
                             \hline
   \multirow{2}{*}{$\beta$}  & 180 ($c$) & 0.62 & 0.289 \\
                             & 225 	& 1.0 & 0.341\\
                             \hline
   \multirow{2}{*}{$\gamma$} & 45 	& 0.46 & 0.372\\
                             & 90 ($b$)	& 1.7 & 0.299\\
                             & 125  & 0.55 & 0.332 \\
                             \hline
   \multirow{2}{*}{$\delta$} & 45 	& 2.0 & 0.729\\
                             & 90 ($b$) & 1.36 & 0.496 \\

   \hline
   \end{tabular}
\end{table}
}

As seen in Table \ref{tab:effmass}, $m^{*}$ scales with the size of the extremal cross-sectional orbit with the exception of the $\gamma$ pocket, which shows an enhanced effective mass for $H\parallel b$. The presence of open orbits for fields aligned aligned along the crystallographic axes, due to the topography of the Fermi surface is likely to affect the angular dependence of the magnetoresistance and this might explain the butterfly shaped magnetoresistance. As illustrated in Fig. \ref{fig:TorqueEffMass}(d), the calculated average effective mass is enhanced along high symmetry directions due to the presence of open orbits but depends on the position of the Fermi level. For small shifts ranging from -11 to $-22\,\mathrm{meV}$ the cyclotron effective mass shows a maximum along the $b$ and the $c$ axes but a minimum in between. This would result in a lower carrier mobility $\nu = q\tau/m^{*}$ (where $q$ is the charge of the carrier and $\tau$  the average scattering time) and hence in a smaller magnetoresistivity for fields along the crystallographic $b$- and the $c$-axis, when compared to fields oriented in between both axes which yield closed cyclotron orbits.

\section{Summary}

In summary, we found that the geometry of the Fermi surface of NbIrTe$_{4}$ obtained from the DFT calculations and from quantum oscillation measurements are in rather good agreement, if one considers a small shift in the position of the Fermi energy which does not affect the existence of Weyl type-II nodes. Our overall results support the presence of Weyl points in this material although it displays rather conventional transport properties. Furthermore, our results contrast to our previous studies on $T_d$-MoTe$_2$ \cite{daniel} and WP$_2$ \cite{Rico}, which are also predicted to display a Weyl type-II semimetallic state, but whose experimental Fermi surfaces derived from quantum oscillations can only be captured by DFT after electron and hole bands are independently displaced towards higher and lower energies, respectively. For both compounds this ad-hoc procedure would suppress their Weyl points, in contrast to their robustness in NbIrTe$_4$. Both quantum oscillations and Hall-effect measurements indicate that NbIrTe$_{4}$ is not a compensated semimetal and that the unconventional four-fold anisotropy of its angular magnetoresistivity is governed by the topography of its Fermi surface and related anisotropy in effective masses. The good agreement between bulk measurements and band structure calculations imply that the electronic properties of NbIrTe$_{4}$, and hence its topological character, are well-captured by Density Functional Theory calculations. More importantly, since this compound is exfoliable and predicted by DFT to display a quantum-spin-Hall insulator state in the monolayer limit \cite{liu_van_2017}, as recently found for WTe$_2$ \cite{pablo}, our conclusions convey that it would be important to explore edge conduction and the effect of a gate voltage in heterostructures containing monolayers of NbIrTe$_{4}$.


\section{Acknowledgments}

We thank S. Sur and V. Quito for helpful discussions. This work was supported by DOE-BES through award DE-SC0002613. JYC acknowledge NSF DMR-1700030 for partial support.
The NHMFL is supported by NSF through NSF-DMR-1644779 and the State of Florida.

\end{document}